\documentclass{article}
\usepackage{spconf,amsmath,graphicx,hyperref}

\usepackage{booktabs}
\usepackage{graphicx}
\title{Feedback-Guided DNN-Based Controller Fusion for Robust Fixed-Parameter Active Noise Control}

\name{Lu Bai$^{1,2}$, Yiming He$^{1}$, Xiaofeng Nan$^{1}$, Kai Chen$^{1}$, Jing Lu$^{1,2,*}$}

\address{
	$^{1}$Key Lab of Modern Acoustics, Institute of Acoustics, Nanjing University, Nanjing, China\\
	$^{2}$NJU-Horizon Intelligent Audio Laboratory, Nanjing Institute of Advanced Artificial \\ Intelligence, Nanjing, China\\
	\{lubai, ymhe, xfnan\}@smail.nju.edu.cn,
	\{kaichen, lujing\}@nju.edu.cn
}

\begin{document}
\ninept
\maketitle
\let\thefootnote\relax
\footnotetext{$^{*}$Corresponding author.}
\begin{abstract}
	In active noise control (ANC) systems, adaptive approaches may suffer from instability or divergence, limiting their practical deployment. Consequently, fixed-parameter controllers are widely adopted, but their performance degrades under varying noise characteristics and acoustic path conditions. This paper proposes a feedback-guided DNN-based controller fusion framework for robust fixed-parameter ANC. The proposed method combines a causal WaveNet controller with a feedback-guided mixture-of-experts (MoE) module, where a gating network estimates the weights of multiple pre-trained FIR experts according to the current acoustic condition. The proposed approach improves robustness to varying acoustic conditions without online parameter updating. Furthermore, the model is fully causal and supports sample-wise streaming inference, with computational costs evenly distributed across sampling points to reduce peak computational load. Experimental results on headphone ANC demonstrate substantial low-frequency noise reduction with negligible noise amplification over 1--8~kHz.
\end{abstract}

\vspace{-1.25em}
\section{Introduction}
\vspace{-0.75em}
\label{sec:intro}
Active noise control (ANC) mitigates unwanted noise by generating anti-noise signals that produce destructive interference in the target region \cite{kuo1999active}. With the growing demand for acoustic comfort and quieter environments, ANC has been widely adopted in applications such as headphones \cite{1597204, shen2022adaptive}, automotive cabins \cite{lian2024online, he2026neural}, aircraft \cite{elliot1990flight}, and other fields \cite{wang2022improving}. 

Conventional ANC systems widely employ adaptive filters as controllers, which update their parameters according to the measured error signals to accommodate changing acoustic environments. Although numerous adaptive algorithms have been developed to improve convergence speed and computational efficiency \cite{lian2024online, he2025modified}, online adaptation still requires a convergence process and may raise stability concerns under secondary-path modeling errors or rapidly varying acoustic conditions \cite{lu2021survey}. To improve the response speed and stability, fixed-parameter ANC controllers are widely adopted in systems for headphones \cite{fabry2019acoustic}, windows \cite{bhan2016feasibility} and rooms \cite{haasjes2024efficient}. However, their performance can degrade when the actual noise characteristics or acoustic path conditions differ from those considered during controller optimization \cite{luo2025deep}.

Recently, deep learning has been introduced into ANC systems to develop fixed-parameter controllers with stronger modeling capabilities \cite{zhang2021deep, zhang2023low, bai2025wavenet, zhang2026causal}. Early DNN-based ANC methods \cite{zhang2021deep, zhang2023low} adopted deep neural networks (DNNs) as ANC controllers, but their non-causal architectures and limited noise reduction performance hindered practical deployment. The fully causal WaveNet-VNN \cite{bai2025wavenet} further improved DNN-based ANC controller by integrating the temporal modeling capability of WaveNet with the nonlinear modeling capability of Volterra neural networks (VNNs), achieving superior noise reduction performance under various noise types and acoustic conditions. More recently, the causal MVNet \cite{zhang2026causal} incorporated time- and frequency-domain information to further enhance ANC performance in road noise control. Nevertheless, existing DNN-based ANC controllers are generally trained offline and their performance remains dependent on the noise characteristics and acoustic conditions included in the training data. When the test conditions differ significantly from the training distribution, their noise reduction performance may degrade substantially \cite{bai2026adaptive}.

To improve the robustness of fixed-parameter ANC under diverse noise conditions, Luo et al. proposed a series of Selective Fixed-Filter ANC (SFANC) \cite{luo2022hybrid, luo2024real, luo2025frequency} and Generative Fixed-Filter ANC (GFANC) \cite{luo2023deep, luo2023delayless, luo2024gfanc, luo2025deep} frameworks. These approaches employ multiple pre-trained control filters and utilize deep neural networks to select or generate appropriate control filters according to the characteristics of the input noise, thereby improving the robustness of fixed-parameter ANC without online controller adaptation. However, their controller selection or generation is primarily based on reference-side noise features, which characterize the input noise but do not directly reflect the actual control outcome under the current acoustic system. Therefore, their ability to respond to acoustic-path mismatch may remain limited.

\begin{figure*}[t]
	\centering
	\includegraphics[width=0.95\textwidth]{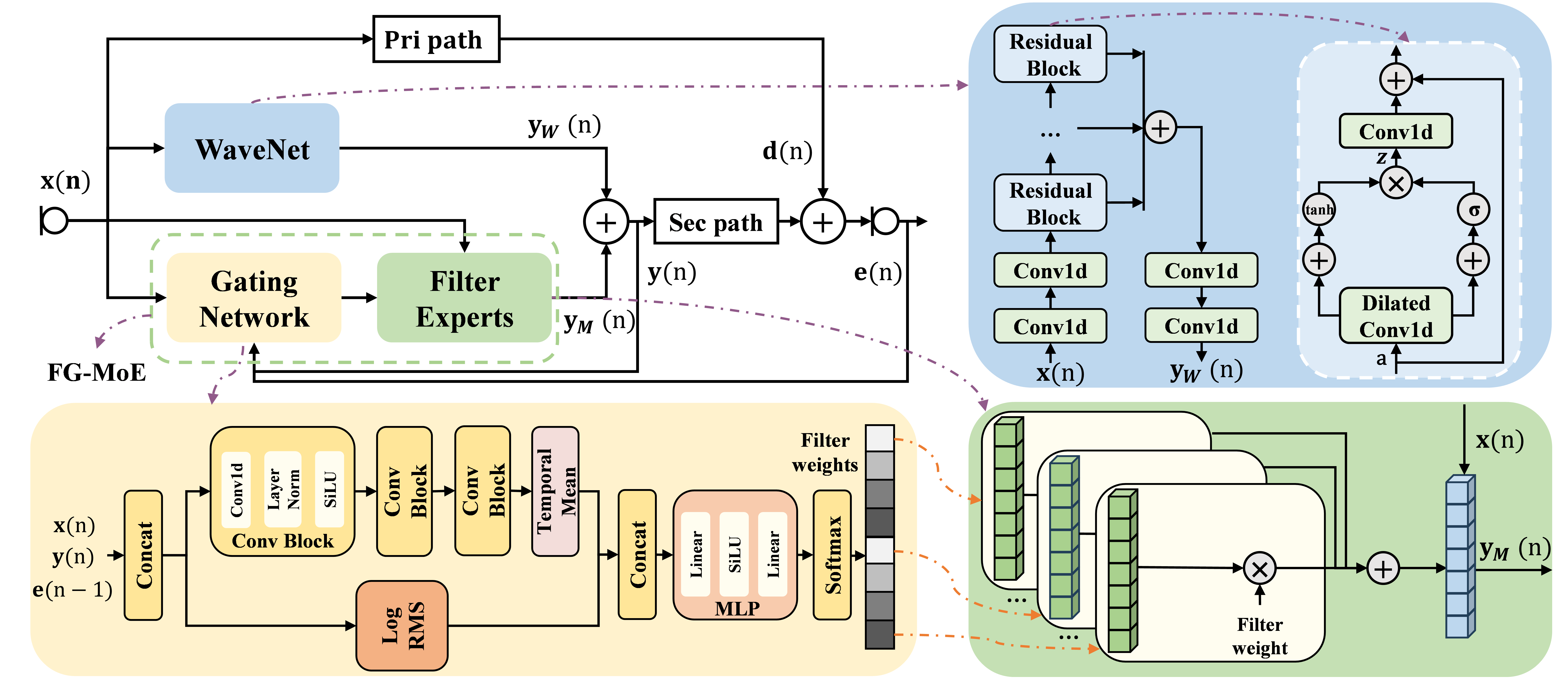}
	\caption{Overall architecture of the proposed feedforward--feedback hybrid ANC system, including the feedforward WaveNet controller branch, the gating network, and the filter experts.}
	\label{fig:framework}
	\vspace{-1em}
\end{figure*}
To address these limitations, this paper proposes a feedback-guided DNN-based controller fusion framework for robust fixed-parameter ANC. Unlike existing SFANC and GFANC approaches that determine the controller primarily from reference-side features, the proposed framework further exploits the control and residual-error signals to characterize the actual control outcome under the current acoustic system. Specifically, a causal WaveNet provides a general control component, while a feedback-guided gating network dynamically combines multiple pre-trained FIR experts for condition-specific correction. A frequency-aware loss jointly promotes one-third-octave-band noise reduction and suppresses high-frequency rebound, while a staged training strategy progressively optimizes the WaveNet, path-specific FIR experts, and gating network. In addition, a fully causal streaming implementation distributes the computational cost across sampling points to reduce the peak computational load. Numerical simulations on headphone ANC under various noise and acoustic-path conditions demonstrate the effectiveness of the proposed framework.

\vspace{-1em}
\section{THEORY}
\vspace{-0.75em}
\label{sec:THEORY}
Fig.~\ref{fig:framework} illustrates the overall architecture of the proposed method. The proposed method replaces the conventional control filter with a DNN-based controller. Let $\mathbf{p}(n)$ and $\mathbf{s}(n)$ denote the primary and secondary paths, respectively. The reference signal $\mathbf{x}(n)$ is processed by the proposed DNN-based controller to generate the control signal $\mathbf{y}(n)$. The primary noise signal $\mathbf{d}(n)$ and error signal $\mathbf{e}(n)$ are expressed as
\begin{equation}
	\label{eq:anc_system}
	\begin{aligned}
		\mathbf{d}(n)&=\mathbf{p}(n)*\mathbf{x}(n),\\
		\mathbf{e}(n)&=\mathbf{d}(n)+\mathbf{s}(n)*\mathbf{y}(n),
	\end{aligned}
\end{equation}
where $*$ denotes linear convolution.

Specifically, the proposed DNN-based controller consists of two parallel branches: a feedforward WaveNet controller branch and a feedback-guided MoE branch. The outputs of the two branches are weighted and summed to generate the final anti-noise signal, which is formulated as
\begin{equation}
	\label{eq:control_signal}
	\mathbf{y}(n)=\alpha \mathbf{y_{\mathrm{W}}}(n)+(1-\alpha)\mathbf{y_{\mathrm{M}}}(n),
\end{equation}
where $\mathbf{y_{W}}(n)$ and $\mathbf{y_{M}}(n)$ denote the outputs of the feedforward WaveNet controller branch and the feedback-guided MoE branch, respectively, and $\alpha$ represents the fusion coefficient between the two branches.

\vspace{-1em}
\subsection{Feedforward WaveNet Controller}
\vspace{-0.5em}
The feedforward controller branch directly maps the reference signal sequence to the control signal sequence through a DNN. This branch exploits the nonlinear modeling capability of neural networks and provides a stable baseline controller. Its parameters are optimized using the entire training dataset, resulting in an averaged optimal solution over different acoustic conditions. Since the controller parameters remain fixed during operation and are independent of the current reference or error signals, this branch provides a stable performance baseline for the proposed adaptive fusion framework.

In this paper, we adopt WaveNet as the feedforward controller branch, as previous studies have demonstrated its effectiveness in various ANC tasks \cite{bai2025wavenet}. The VNN module is not incorporated because the nonlinear effects in the considered task are relatively limited, while avoiding the additional computational cost introduced by the higher-order nonlinear modeling. For scenarios involving stronger nonlinear distortions, this branch can be readily replaced by a more expressive architecture, such as WaveNet-VNN or other suitable nonlinear neural controllers.

As shown in Fig.~\ref{fig:framework}, the WaveNet controller takes the reference signal sequence $\mathbf{x}(n)$ as input and directly generates the control signal sequence $\mathbf{y}_{\mathrm{W}}(n)$. Two one-dimensional convolutional (Conv1d) layers are applied after the input and before the output to adjust the number of channels. The backbone consists of a stack of residual blocks, each consisting of a dilated one-dimensional convolution (Dilated Conv1d) layer, a gated unit, and a Conv1d layer. The dilated convolution enlarges the receptive field without significantly increasing the number of parameters, enabling the network to capture long-term temporal dependencies in the reference signal. Within each residual block, the output of the Dilated Conv1d is further processed by the gated activation unit to enhance the nonlinear modeling capability of the network. Specifically, the gated activation unit is formulated as
\begin{equation}
	\label{equation1}
	\mathbf{z} = 
	\tanh \left( W_{f,k} * \mathbf{a}  \right)
	\odot
	\sigma \left( W_{g,k} * \mathbf{a} \right),
\end{equation}
where $\ast$ denotes the convolution operation, $\odot$ represents the element-wise multiplication operator, $\sigma(\cdot)$ is the sigmoid function, $k$ is the layer index, $f$ and $g$ correspond to the filter and gate, respectively, and $W$ is a learnable convolution kernel.

\vspace{-1em}
\subsection{Feedback-Guided MoE}
\vspace{-0.5em}
As shown in Fig.~\ref{fig:framework}, the feedback-guided MoE module consists of a gating network and multiple filter experts. The gating network estimates the fusion weights of different experts based on the current acoustic condition, while the filter experts consist of multiple pre-trained finite impulse response (FIR) filter controllers with different coefficients. The coefficients of these experts are dynamically weighted and combined to obtain the MoE controller:
\begin{equation}
	\label{eq:moe_filter}
\mathbf{w}_{\mathrm{M}}(n)=\sum_{i=1}^{N}\beta_i(n-1)\mathbf{w}_i,
\end{equation}
where $N$ represents the total number of experts, $\mathbf{w}_i$ denotes the coefficients of the $i$-th filter expert, and $\beta_i(n-1)$ represents the corresponding fusion weight estimated at the previous sampling point. 
The output of the MoE controller is then obtained as
\begin{equation}
	\label{eq:moe_output}
	\mathbf{y_{\mathrm{M}}}(n)=\mathbf{w}_{\mathrm{M}}(n)*\mathbf{x}(n).
\end{equation}

\vspace{-1em}
\subsubsection{Gating Network}
\vspace{-0.5em}
The gating network aims to estimate the fusion weights of different FIR controller experts according to the current acoustic condition. As shown in Fig.~\ref{fig:framework}, the inputs of the gating network include the reference signal $\mathbf{x}(n)$, control signal $\mathbf{y}(n)$, and error signal $\mathbf{e}(n-1)$. At sampling point $n$, the current control signal $\mathbf{y}(n)$ is first generated using the fusion weights $\boldsymbol{\beta}(n-1)$ estimated at the previous sampling point. The gating network then uses $\mathbf{x}(n)$, $\mathbf{y}(n)$, and $\mathbf{e}(n-1)$ to update the fusion weights to $\boldsymbol{\beta}(n)$, which are applied to generate the control signal at sampling point $n+1$. Therefore, no current output depends on gating weights computed from that same output, and the proposed feedback-guided fusion remains causal. These signals are first concatenated along the channel dimension and then processed by two parallel feature extraction branches. The one-sample delay of the error signal is introduced because $\mathbf{e}(n)$ becomes available only after $\mathbf{y}(n)$ has propagated through the secondary path. Although the error signal is not exactly aligned with the reference and control signals, the introduced delay has negligible influence on the performance, as verified by experimental results.

The first branch employs three cascaded convolutional blocks to extract temporal features from the input sequences. Each convolutional block consists of a one-dimensional convolution (Conv1d) layer, a layer normalization (Layer Norm) operation, and a SiLU activation function. After the convolutional blocks, a temporal mean operation is applied to obtain a global temporal representation. The second branch calculates the logarithmic root mean square (Log RMS) features to characterize the amplitude statistics of the input signals. The features extracted from the two branches are then concatenated and fed into an MLP consisting of two linear layers and a SiLU activation function. Finally, a Softmax layer is applied to generate the fusion weights of different FIR controller experts.

\vspace{-1em}
\subsubsection{Filter Experts}
\vspace{-0.5em}
The filter experts consist of multiple pre-trained FIR filter controllers with different coefficients. In this work, each filter expert is directly implemented using a Conv1d layer with a kernel size of $2048$. This implementation is equivalent to a 2048-tap FIR filter without coefficient flipping. The coefficients of these filter experts are dynamically weighted according to the fusion weights estimated by the gating network, generating the final MoE controller.

\vspace{-1em}
\subsection{Loss Function and Training Strategy}
\vspace{-0.5em}
We employ a frequency-aware ANC loss based on one-third-octave-band analysis. The noise-reduction metric is the equally weighted mean noise reduction over 50~Hz--5~kHz, while the rebound metric is the largest noise amplification over 1~kHz--8~kHz and is set to zero when no band is amplified. Accordingly, $\mathcal{L}_{\mathrm{NR}}$ is defined as the negative noise-reduction metric. For a more conservative constraint, $\mathcal{L}_{\mathrm{RB}}$ extends the upper frequency of the rebound metric to 16~kHz. Both terms are calculated from the disturbance $\mathbf{d}$ and residual $\mathbf{e}$ using an 8192-point STFT with a Hann window and a hop size of 2048. A broadband NMSE term is further included to constrain the overall residual energy:
\begin{equation}
	\mathcal{L}_{\mathrm{NMSE}}=10\log_{10}\frac{\sum_n e^2(n)}{\sum_n d^2(n)}.
\end{equation}
The batch-averaged ANC objective is
\begin{equation}
	\mathcal{L}_{\mathrm{ANC}}=\mathcal{L}_{\mathrm{NR}}+\lambda\mathcal{L}_{\mathrm{RB}}+\mathcal{L}_{\mathrm{NMSE}},
\end{equation}
where $\lambda$ is a weighting coefficient that balances noise reduction and rebound suppression. 

Training proceeds in three stages. First, the WaveNet controller is trained on all training conditions for 180 epochs. Second, each FIR expert is trained independently for 180 epochs using the data associated with its acoustic path; these two pre-training stages can be performed in parallel. Third, the pre-trained WaveNet and FIR experts are frozen, and only the gating network is optimized for 100 epochs. For this stage, a cross-entropy loss $\mathcal{L}_{\mathrm{cls}}$ supervised by the acoustic-path labels, with a label-smoothing factor of 0.05, is added as
\begin{equation} 
	\mathcal{L}_{\mathrm{total}}
	=\mathcal{L}_{\mathrm{ANC}}
	+\gamma\mathcal{L}_{\mathrm{cls}}, 
\end{equation} 
where $\gamma$ is a weighting coefficient that balances the contributions of the ANC loss and the classification loss. Here, $\mathcal{L}_{\mathrm{cls}}$ establishes the correspondence between acoustic paths and their associated FIR experts, while $\mathcal{L}_{\mathrm{ANC}}$ refines the soft fusion weights according to the resulting ANC performance. 

\vspace{-1em}
\section{NUMERICAL SIMULATIONS}
\vspace{-0.75em}
\label{sec:numerical_simulations}

\subsection{Simulation Setup}
\vspace{-0.5em}
The simulations are conducted using the headphone ANC dataset provided by the second track of the CCF Audio and Acoustic Technology Challenge (CCF-AATC), available at \url{https://ccf-aatc.org.cn/}. This dataset is designed for evaluating headphone ANC systems under diverse noise and acoustic conditions. Specifically, the dataset contains eight noise scenarios, 80 primary noise recordings, and ten sets of secondary paths. Each noise scenario includes a 30-min reference signal, resulting in eight 30-min reference recordings. The primary noise dataset consists of 80 30-min recordings corresponding to different path conditions, while the ten secondary paths are provided to model the acoustic transfer responses from the loudspeaker to the error microphone. All data are subsampled to 48 kHz.

The first eight paths and six noise scenarios form the training set, with one FIR expert trained for each path. Primary noise recordings and the corresponding secondary paths are randomly sampled during training. The remaining two paths and two noise scenarios are reserved to evaluate robustness to unseen conditions.

\vspace{-1em}
\subsection{Simulation Results}
\vspace{-0.5em}
We first evaluate the noise reduction performance under unseen noise conditions with seen and unseen acoustic paths. The official CCF 2026 ANC baseline model provided by the challenge organizers (\url{https://github.com/CCF2026ANC/CCF_DEEPANC_2026}) and a standalone feedforward WaveNet branch are adopted as comparison methods. The official CCF 2026 ANC baseline model provided by the challenge organizers is trained using the original NMSE loss and training strategy provided by the challenge. In contrast, the feedforward WaveNet branch and the proposed hybrid ANC network are trained using the optimized loss function and training strategy introduced in this work. The numbers of parameters of the official baseline, the feedforward WaveNet branch, and the proposed hybrid ANC network are 42.76k, 10.08k, and 28.57k, respectively. Their corresponding computational costs are 2.04~GMac/s, 483.84~MMac/s, and 672.83~MMac/s, respectively. 

\begin{figure}[t]
	\centering
	\includegraphics[width=\linewidth]{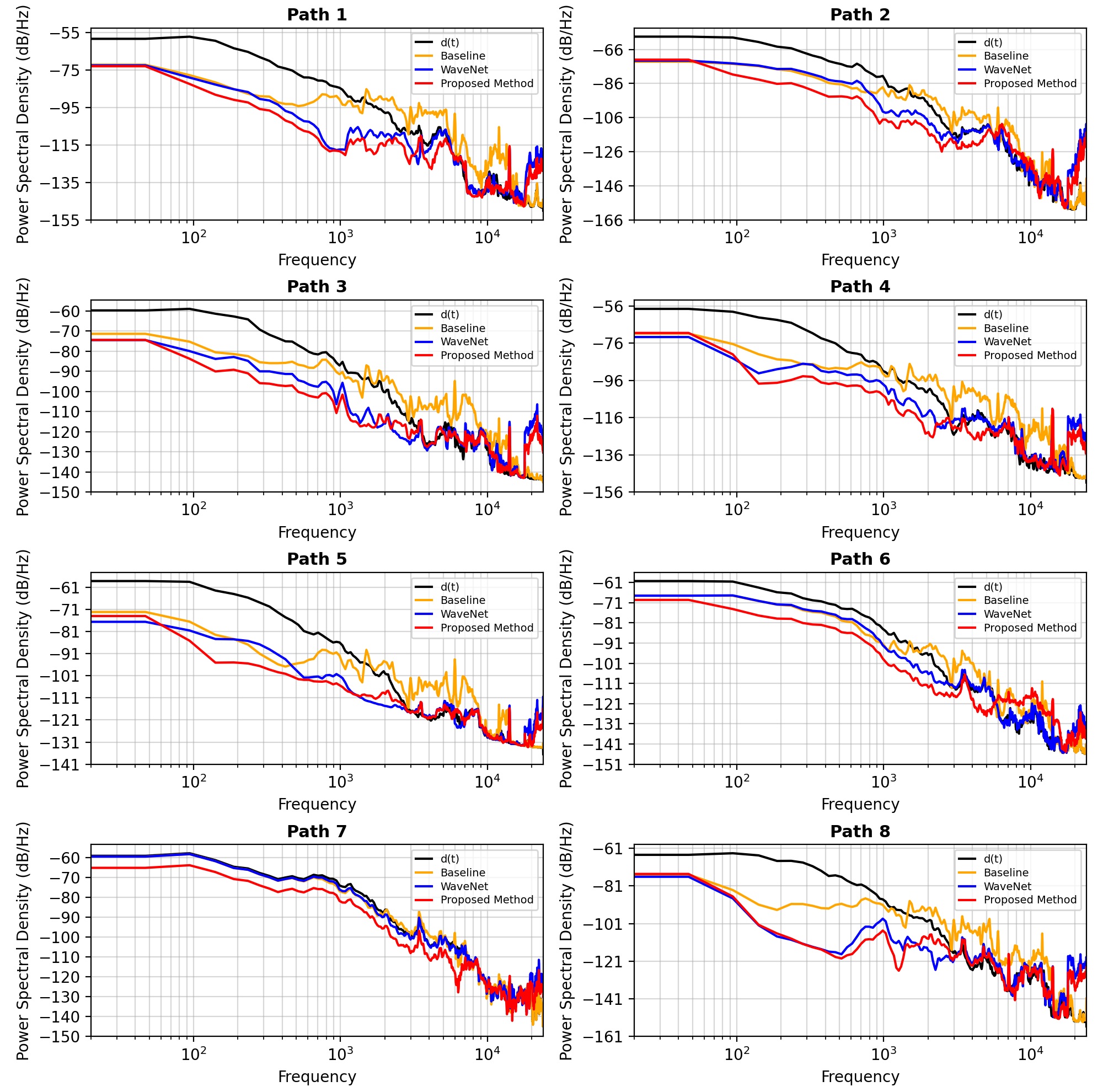}
	\caption{Noise reduction spectra of different ANC methods under seen acoustic paths with unseen noise.}
	\label{fig:combined_seen_paths_freq}
\end{figure}

Fig.~\ref{fig:combined_seen_paths_freq} presents the noise reduction spectra of different ANC methods under seen acoustic paths with unseen noise. Although the standalone feedforward WaveNet branch has fewer parameters and lower computational cost than the official baseline, it achieves superior performance when trained with the optimized loss function and training strategy proposed in this work. However, the standalone feedforward WaveNet controller still exhibits limited robustness to acoustic path variations. In particular, under the seventh seen path, it provides almost no noise reduction because this path has a significantly different acoustic response from the other training paths. Since the feedforward WaveNet controller is trained with a single set of parameters shared across all acoustic paths, it can only approximate an average optimal solution over multiple path conditions. Consequently, its performance is significantly degraded for acoustic paths with large acoustic differences from the others. 

\begin{figure}[!t]
	\centering
	\includegraphics[width=\linewidth]{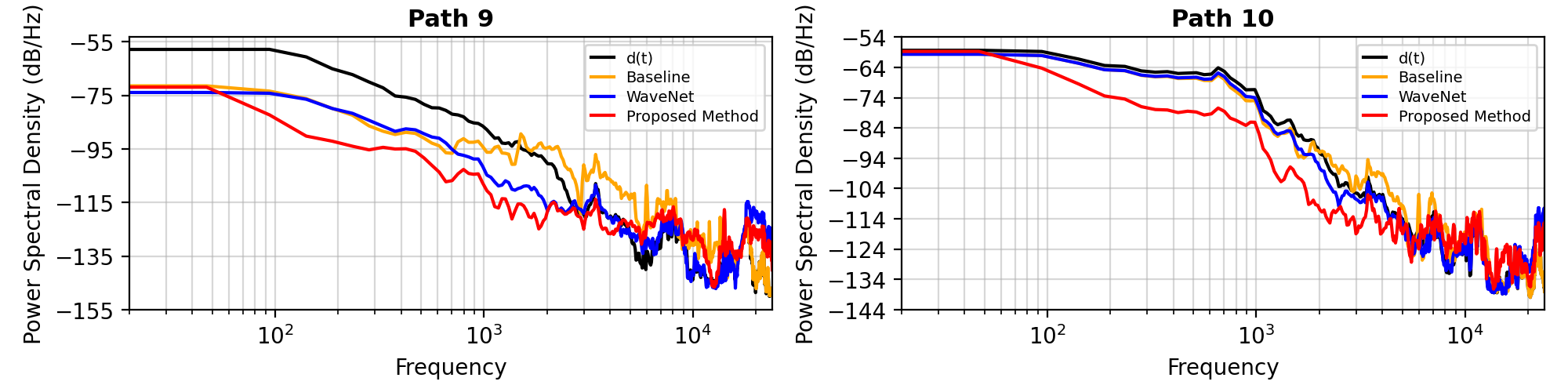}
	\caption{Noise reduction spectra of different ANC methods under unseen acoustic paths with unseen noise.}
	\label{fig:combined_unseen_paths_freq}
	\vspace{-1em}
\end{figure}

The proposed method further integrates the feedback-guided MoE branch, which achieves substantially improved low-frequency noise reduction compared with the standalone feedforward WaveNet branch, particularly under the seventh acoustic path. However, the feedback-guided MoE branch may introduce larger high-frequency amplification under certain acoustic paths. For example, under the sixth acoustic path, its high-frequency noise amplification is slightly higher than that of the standalone WaveNet branch. By combining the feedforward WaveNet branch and the feedback-guided MoE branch, the proposed hybrid architecture effectively exploits their complementary advantages, improving low-frequency noise reduction while alleviating high-frequency amplification. Fig.~\ref{fig:combined_unseen_paths_freq} presents the noise reduction spectra of different ANC methods under unseen acoustic paths. It can be observed that the proposed method achieves superior noise reduction performance on both unseen acoustic paths, demonstrating its robustness to acoustic path variations.

\begin{figure}[t]
	\centering
	\includegraphics[width=0.9\linewidth]{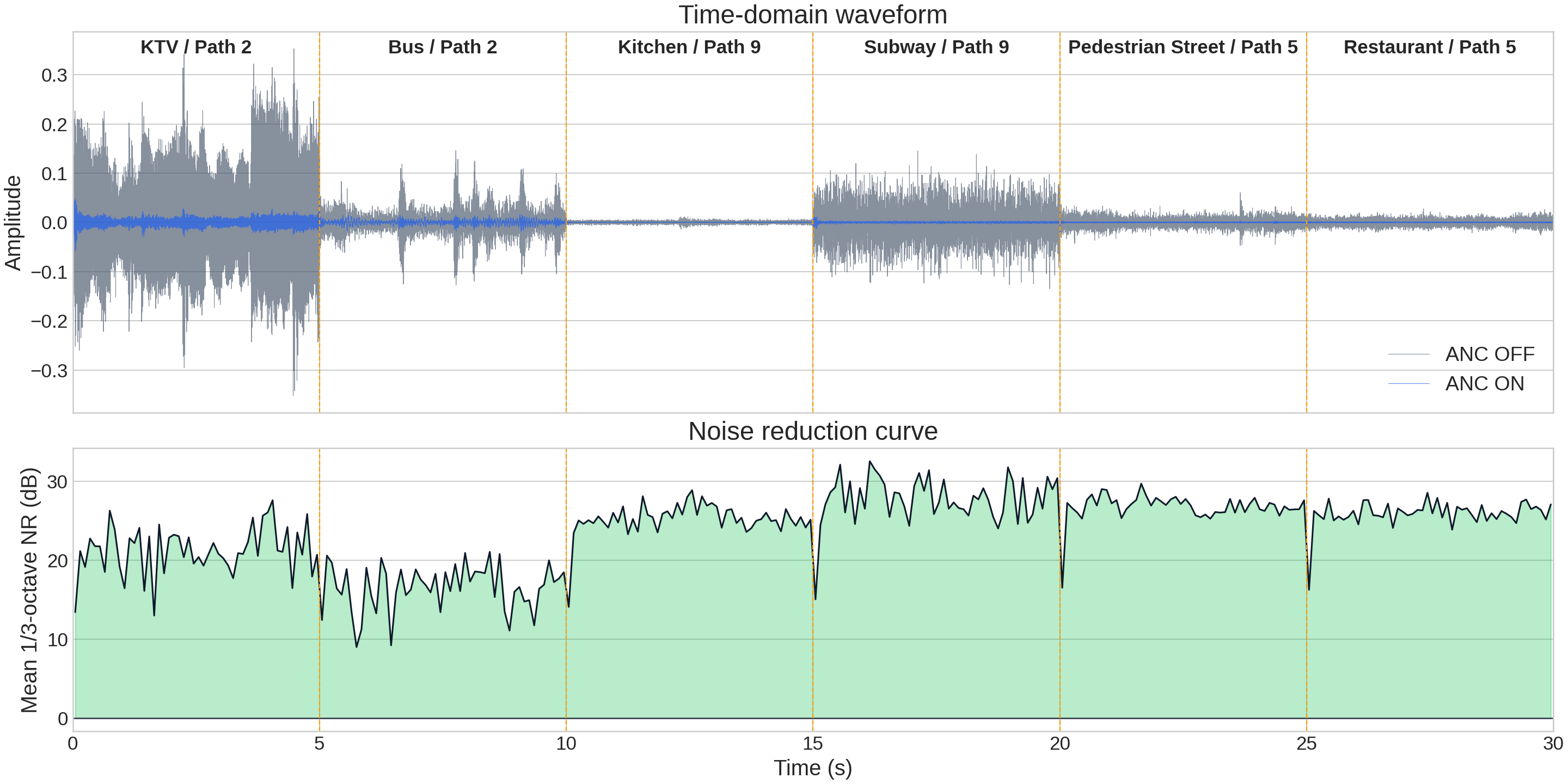}
	\caption{Time-domain noise reduction performance of the proposed ANC model under different acoustic paths and noise conditions.}
	\label{fig:random_30s_time_domain}
	\vspace{-1em}
\end{figure}

\begin{figure}[t]
	\centering
	\includegraphics[width=0.9\linewidth]{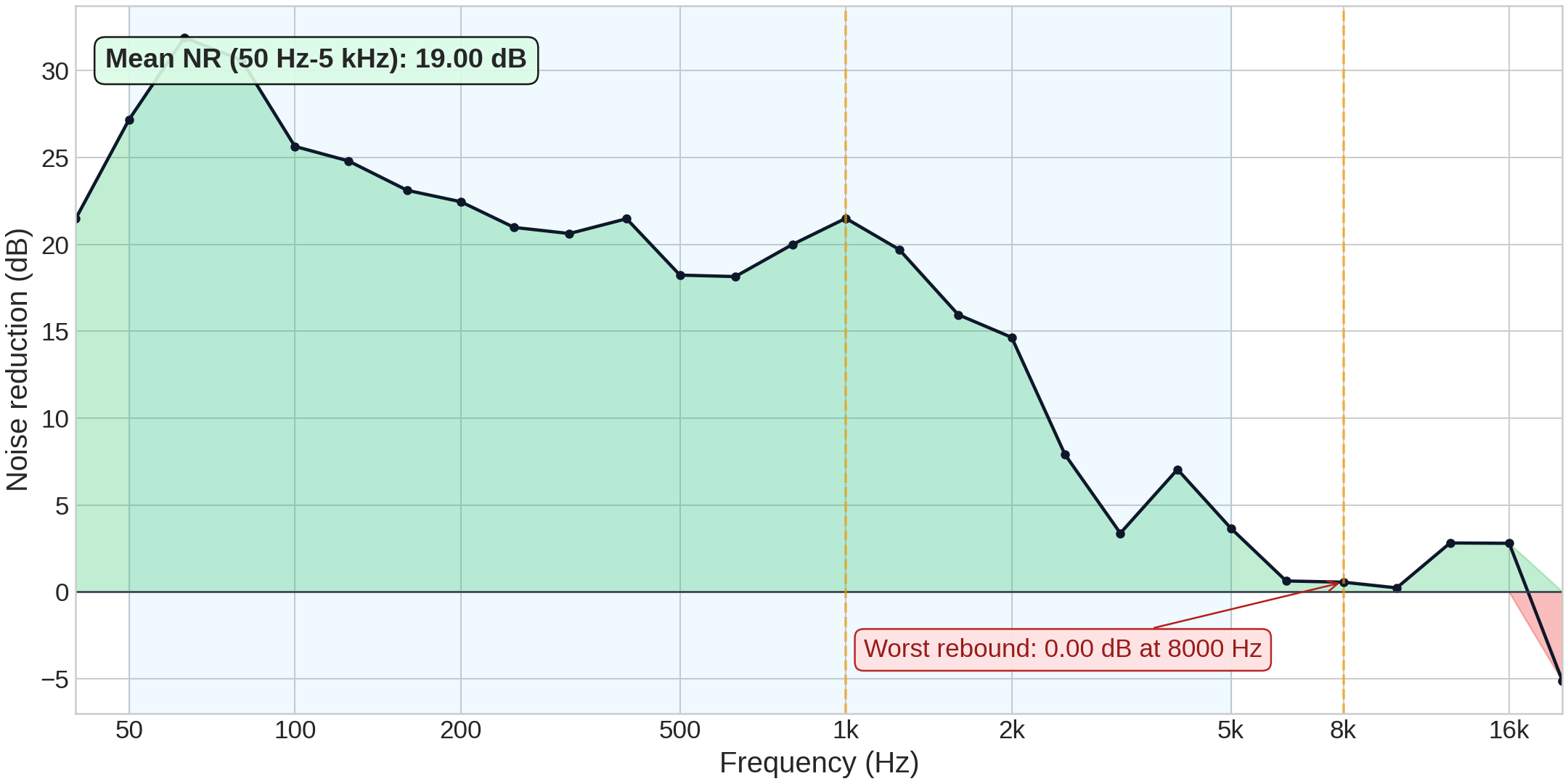}
	\caption{Third-octave band noise reduction performance of the proposed ANC model.}
	\label{fig:random_30s_third_octave}
	\vspace{-1em}
\end{figure}

Furthermore, we train a ten-expert model on the complete dataset, with one FIR expert for each path. It is evaluated on three randomly selected paths and two noise conditions per path, forming six 5-s cases (30~s in total). Fig.~\ref{fig:random_30s_time_domain} shows that the model maintains stable noise reduction across condition switches without convergence time. The corresponding third-octave-band result in Fig.~\ref{fig:random_30s_third_octave} shows an average noise reduction of 19.00~dB from 50~Hz to 5~kHz with negligible noise amplification over 1--8~kHz.

For the ten-expert model, we further develop a fully streaming implementation with peak-MAC optimization. It contains 32.69k parameters, compared with 28.57k for the eight-expert model, due to two additional 2048-tap FIR experts and their gating outputs. Its cost is 672.93~MMac/s, versus 672.83~MMac/s for the eight-expert model. Compared with standard streaming inference, peak-MAC optimization reduces the peak MAC from 34.62k to 14.15k while leaving the overall complexity essentially unchanged.

\vspace{-1em}
\section{CONCLUSION}
\vspace{-0.75em}
\label{sec:CONCLUSION}

This paper proposes a feedback-guided DNN-based controller fusion framework for robust fixed-parameter ANC. The proposed framework integrates a causal WaveNet controller with a feedback-guided MoE module, where residual-error and control signals are utilized to dynamically fuse multiple pre-trained FIR experts according to the current acoustic condition. By incorporating the effects of acoustic transfer paths into controller fusion, the proposed method improves the robustness of fixed-parameter ANC without online parameter updating. Furthermore, a fully causal sample-wise streaming implementation is developed to enable efficient sample-wise inference, and the computational cost is evenly distributed across sampling points to reduce the peak computational load. Numerical simulations on headphone ANC demonstrate that the proposed framework consistently outperforms baseline methods under different noise conditions and acoustic paths, achieving substantial low-frequency noise reduction with negligible noise amplification over 1--8~kHz.

\newpage
\vfill\pagebreak


\bibliographystyle{IEEEbib}
\bibliography{strings,refs}

\end{document}